\documentclass[
    aps,prb,twocolumn,
	groupedaddress,superscriptaddress,
	amsfonts,amssymb,amsmath,
	citeautoscript,longbibliography,
	letterpaper
	]{revtex4-2}

\usepackage[utf8]{inputenc}
\usepackage[english]{babel}

\usepackage{microtype} 
\usepackage{xspace} 

\usepackage{txfonts}  
\usepackage{txfontsb} 

\usepackage{bm} 

\usepackage[usenames,dvipsnames]{xcolor}
\usepackage[]{graphicx} 
\graphicspath{{figs/}}

\usepackage[]{booktabs}
\usepackage{array}
\usepackage{layouts}
\usepackage{multirow}

\usepackage{enumerate}
\usepackage[inline]{enumitem}

\usepackage{xr}
\makeatletter
\newcommand*{\addFileDependency}[1]{
  \typeout{(#1)}
  \@addtofilelist{#1}
  \IfFileExists{#1}{}{\typeout{No file #1.}}
}
\makeatother

\usepackage{hyperref}
\hypersetup{colorlinks,
	linkcolor={blue!75!black!80!yellow},
	citecolor={blue!75!black!80!yellow},
	urlcolor={blue!75!black!80!yellow}
}

\usepackage[capitalize,nameinlink]{cleveref}

\crefname{subequations}{Eqs.}{Eqs.} 
\Crefname{subequations}{Eqs.}{Eqs.}
\crefformat{subequations}{#2Eqs.~(#1)#3}
\Crefformat{subequations}{#2Eqs.~(#1)#3}
\crefname{page}{p.}{p.} 

\usepackage{placeins}

\usepackage{siunitx}
\DeclareSIUnit[number-unit-product = ]\percent{\char`\%} 

\usepackage[centering,hmargin=18mm,tmargin=27mm,bmargin=22mm]{geometry}

\usepackage{soul}

\usepackage{textcomp} 
\usepackage{xifthen}
\usepackage{etoolbox}
\newboolean{togglecomments}
\newboolean{toggletodos}
\newboolean{togglechanges}

\setboolean{togglecomments}{true}
\setboolean{toggletodos}{true}
\setboolean{togglechanges}{false} 

\newcommand{\textblacksquare}{$\blacksquare$}
\newcommand{\todo}[1]{\ifbool{toggletodos}%
	{\textcolor{green!60!black}{\small\textsf{{}\textsuperscript{\textsc{\textsf{todo}}}}[\ignorespaces#1]}} 
	{}}     
\newcommand{\comment}[2]{\ifbool{togglecomments}%
		{\textcolor{blue!70!black}{\small\sf\textsuperscript{\textsc{\textsf{\ignorespaces#1}}}[\ignorespaces#2]}} 
		{}}     
\newcommand{\swap}[2]{\ifbool{togglechanges}
	{\ignorespaces#2}  
	{\textcolor{red!70!black}{[\ignorespaces#1]}\textrightarrow{}\textcolor{green!50!black}{[\ignorespaces#2]}}}
\newcommand{\remove}[1]{\ifbool{togglechanges}
	{}    
	{\textcolor{red!70!black}{\ignorespaces#1}}}
\newcommand{\inset}[1]{\ifbool{togglechanges}
	{\ignorespaces#1}  
	{\textcolor{green!50!black}{\ignorespaces#1}}}

\newcommand{\citeremind}[1]{%
	[\textcolor{blue!75!black!80!yellow}{\textblacksquare%
		\ifthenelse{\isempty{#1}}{}{\textsuperscript{\tiny\textsf{\ignorespaces#1}}}%
	}]\xspace}

\newcommand{\ie}{i.e.,\@\xspace} 

\newcommand{\eg}{e.g.,\@\xspace}

\newcommand{\appropto}{\mathrel{\vcenter{
			\offinterlineskip\halign{\hfil$##$\cr
				\propto\cr\noalign{\kern.2pt}\sim\cr\noalign{\kern-2.5pt}}}}}

\DeclareFontFamily{U}{mathx}{\hyphenchar\font45}
\DeclareFontShape{U}{mathx}{m}{n}{<5> <6> <7> <8> <9> <10>
                                  <10.95> <12> <14.4> <17.28> <20.74> <24.88>
                                  mathx10}{}
\DeclareSymbolFont{mathx}{U}{mathx}{m}{n}
\DeclareFontSubstitution{U}{mathx}{m}{n}

\makeatletter
\newcommand{\raisemath}[1]{\mathpalette{\raisem@th{#1}}}
\newcommand{\raisem@th}[3]{\raisebox{#1}{$#2#3$}}
\makeatother

\renewcommand{\paragraph}[1]{\vskip .5ex\noindent\emph{#1.}---\ignorespaces}

\usepackage[eulergreek]{sansmath}
\makeatletter
\renewcommand\@make@capt@title[2]{%
    \@ifx@empty\float@link{\@firstofone}{\expandafter\href\expandafter{\float@link}}%
    \sisetup{math-sf=\textsf}%
    \sansmath\sffamily\textbf{#1\@caption@fignum@sep}#2 
}%

\makeatother

\usepackage{physics} 

\graphicspath{{figures/}}

\setboolean{togglecomments}{false}
\setboolean{toggletodos}{true}
\setboolean{togglechanges}{false} 

\begin{document}
\title{Weyl points on non-orientable manifolds}

\author{Andr\'e Grossi Fonseca}
\email{agfons@mit.edu}
\affiliation{Department of Physics, Massachusetts Institute of Technology, Cambridge, Massachusetts~02139, USA}
\author{Sachin Vaidya}
\email{svaidya1@mit.edu}
\thanks{A.G.F.\ and S.V.\ contributed equally}
\affiliation{Department of Physics, Massachusetts Institute of Technology, Cambridge, Massachusetts~02139, USA}
\author{Thomas Christensen}
\affiliation{Department of Electrical and Photonics Engineering, Technical University of Denmark, Kgs.~Lyngby~2800, Denmark}
\author{Mikael C. Rechtsman}
\affiliation{Department of Physics, The Pennsylvania State University, University Park, PA~16802, USA}
\author{Taylor L. Hughes}
\affiliation{Department of Physics and Institute for Condensed Matter Theory,
University of Illinois at Urbana-Champaign, Urbana, IL~61801, USA}
\author{Marin Solja\v ci\'c}
\affiliation{Department of Physics, Massachusetts Institute of Technology, Cambridge, Massachusetts~02139, USA}

\begin{abstract}
Weyl fermions are hypothetical chiral particles that can also manifest as excitations near three-dimensional band crossing points in lattice systems.
These quasiparticles are subject to the Nielsen--Ninomiya ``no-go'' theorem when placed on a lattice, requiring the total chirality across the Brillouin zone to vanish. 
This constraint results from the topology of the (orientable) manifold on which they exist. Here, we ask to what extent the concepts of topology and chirality of Weyl points remain well defined when the underlying manifold is non-orientable. 
We show that the usual notion of chirality becomes ambiguous in this setting, allowing for systems with a non-zero total chirality. 
This circumvention of the Nielsen--Ninomiya theorem stems from a generic discontinuity of the vector field whose zeros are Weyl points.
Furthermore, we discover that Weyl points on non-orientable manifolds carry an additional $\mathbb{Z}_2$ topological invariant which satisfies a different no-go theorem. We implement such Weyl points by imposing a non-symmorphic symmetry in the momentum space of lattice models. 
Finally, we experimentally realize all aspects of their phenomenology in a photonic platform with synthetic momenta. 
Our work highlights the subtle but crucial interplay between the topology of quasiparticles and of their underlying manifold.

\end{abstract}
\maketitle

Weyl fermions are massless particles of definite chirality allowed by the Standard Model of particle physics~\cite{Weyl1929}.
Although they remain elusive as high-energy particles, they can emerge as low-energy excitations in quantum systems~\cite{Xu2015, Lv2015, Armitage2018, wang2021realization}, and their dispersion also appears in certain classical systems~\cite{Lu2015, Li2018, he2020observation, noh2017experimental, vaidya2020observation, jorg2022observation}. 
Such Weyl quasiparticles occur near band degeneracies, known as Weyl points, in the momentum space of three-dimensional lattices, and exhibit a number of unique properties: 
(1)~as monopoles of Berry curvature, they carry an integer topological charge, the first Chern number, the sign of which defines their chirality~\cite{Bradlyn2016, Chang2018}.
They are therefore robust to perturbations, even some that break translational symmetry~\cite{Pixley2018, Fonseca2023, mastropietro2020stability}.
(2)~A bulk--boundary correspondence associates Weyl points with Fermi arcs---dispersive surface states that connect Weyl points of opposite charges~\cite{Wan2011}. 
(3)~In the presence of gauge fields, they can generate a violation of chiral charge conservation, a phenomenon known as the chiral anomaly~\cite{nielsen1983adler}. 

Formulating lattice theories that describe Weyl fermions, or chiral fermions more broadly, imposes global constraints on their chirality.
The most noteworthy example is the Nielsen--Ninomiya theorem, that asserts, under general assumptions of locality, Hermiticity, and translational invariance, that the net chirality of all Weyl fermions vanishes~\cite{Nielsen1981.1, Nielsen1981.2}. In lattice field theories, the Nielsen--Ninomiya theorem generates additional unwanted fermion species, leading to the important problem of fermion doubling~\cite{Luscher2002, Kaplan2012}. 
Additionally, since condensed matter systems are often crystalline, this theorem applies and constrains the total Weyl-point chirality in the Brillouin zone to vanish. 
This, in turn, directly impacts physical observables, e.g.,
the spectrum and dispersion of Fermi arcs~\cite{hasan2017discovery}, the chirality of Landau levels under an applied magnetic field~\cite{nielsen1983adler, ilan2020pseudo}, and electromagnetic responses to circularly polarized light~\cite{Sekh2022, deJuan2017}.
 
Several approaches for circumventing the Nielsen--Ninomiya theorem have been proposed, all of which violate one or more of its assumptions~\cite{Neuberger1998, Hernandez1999, ginsparg1982remnant, kaplan1992method, juvenwang2022, rudner2013anomalous, yu2019circumventing, Ahn2019, melon2020, Wang2017, Konig2023}; however, none have explored the role of the topology of the underlying manifold. 
Indeed, the strong constraint imposed by this no-go theorem ultimately results from the topological properties of the momentum-space fundamental domain, i.e., the toroidal Brillouin zone~\cite{Karsten1981}. 
Here, we demonstrate that the notions of chirality and topology for Weyl points are fundamentally altered on non-orientable momentum-space manifolds. 
We show that, while an absolute notion of chirality of Weyl points becomes inherently ambiguous, a relative chirality still exists, and that non-orientability provides a natural setting for the Nielsen--Ninomiya theorem to be circumvented in an atypical fashion. 
We also show that Weyl points on non-orientable manifolds carry a $\mathbb{Z}_2$ topological charge and have an associated no-go theorem that places global constraints on both the number of Weyl points and their total chirality.
Finally, we experimentally realize such Weyl points in a photonic system endowed with synthetic momenta, paving the way to a wider exploration of the interplay between orientability and chirality. 

\begin{figure}[ht]
    \centerline{%
    \includegraphics[scale=1]{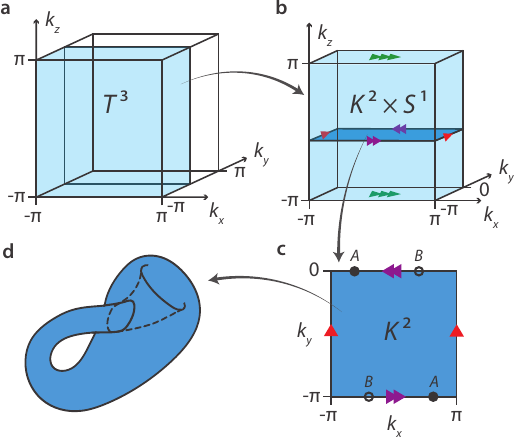}}
    \caption{%
        \textbf{a},~The momentum space of three-dimensional materials is typically represented as a torus, $T^3$. 
        \textbf{b},~In the case of Hamiltonians obeying the symmetry in \cref{eq:H_cons}, the fundamental domain in momentum space takes the form of a non-orientable manifold, $K^2\times S^1$. The arrows indicate boundary identifications. 
        \textbf{c},~Two-dimensional cuts in the $(k_x,k_y)$ plane form a Klein bottle, $K^2$. 
        The Hamiltonian is identical at the pair of points labeled by $A$ and $B$, and similarly along the entire $k_y = -\pi$ and $k_y = 0$ lines.  
        \textbf{d},~$K^2$ can be visualized as a closed manifold created by twisting and gluing one pair of opposing sides of a rectangle and gluing the other pair without a twist (as indicated by the arrows).
        }
    \label{fig:figure1}
\end{figure}

We begin by discussing our scheme for obtaining Hamiltonians of lattice systems with non-orientable momentum-space manifolds.
The Bloch Hamiltonian $H(\mathbf{k})$, in an appropriate basis, is invariant under translations by any reciprocal lattice vector $\mathbf{G}$, i.e., $H(\mathbf{k}) = H(\mathbf{k}+\mathbf{G})$. 
This reflects a redundancy in $\mathbf{k}$-space, since both $\mathbf{k}$ and $\mathbf{k}+\mathbf{G}$ label the same physical momentum point. 
By restricting $\mathbf{k}$ to the set of unique momenta, the Brillouin zone takes the form of a three-dimensional torus, denoted as $T^3$ (\cref{fig:figure1}a).
This toroidal nature of momentum space is unavoidable for a lattice.

Interestingly, under certain circumstances it is possible to subdivide the torus into closed manifolds that are non-orientable~\cite{Chen2022, Wang2023, Zhu2023, zhang2023general, Morimoto2017}. 
This can be achieved, \eg by imposing a momentum-space glide symmetry on the Hamiltonian
\begin{align}
    H(k_x, k_y, k_z) = H(-k_x, k_y+ \pi, k_z),
    \label{eq:H_cons}
\end{align} 
where we have set the lattice constant to unity. This symmetry leads to a further redundancy in $\mathbf{k}$-space, since the Hamiltonian, along with its eigenstates and energy spectrum, is identical at momenta $(k_x, k_y, k_z)$ and $(-k_x, k_y+\pi, k_z)$. 
This symmetry operation  subdivides the torus into two fundamental domains: without loss of generality we select the domain $-\pi \le k_x, k_z < \pi$ and $-\pi \le k_y < 0$ as their representative. 
The non-symmorphic nature of the symmetry in \cref{eq:H_cons} allows for boundary identifications to be made at the $k_y = 0$ and $k_y = -\pi$ planes in a twisted fashion (\cref{fig:figure1}b), resulting in a closed manifold. 
The fundamental domain can consequently be expressed as the direct product of a non-orientable Klein bottle $(K^2)$ in the $(k_x, k_y)$ plane and a circle $(S^1)$ in the $k_z$ direction, \ie $K^2\times S^1$. 
The Klein bottle can be visualized by gluing one pair of opposite sides of a rectangle and twisting and gluing the other pair (\cref{fig:figure1}c). 
\Cref{fig:figure1}d shows an immersion of a Klein bottle in $\mathbb{R}^3$.

We note that the direct equality in \cref{eq:H_cons}, without unitary conjugation of the Hamiltonian, is crucial~\footnote{In \cite{SupplementalMaterial}, we discuss how~\cref{eq:H_cons} is modified if the orbital degrees of freedom are spatially displaced from the origin in the unit cell.}, and distinguishes this symmetry from spatial symmetries, such as rotations, that subdivide the Brillouin zone into identical smaller copies, but which are open manifolds. 
This unitary-free symmetry is analogous to translational symmetry which reduces the domain of the Hamiltonian from $\mathbb{R}^3$ to $T^3$, leading to identical physics at the identified boundaries. 
In further contrast to spatial symmetries, the non-symmorphic nature of this symmetry in momentum space leaves no momentum point invariant.

\begin{figure}
    \centerline{%
    \includegraphics[scale=1]{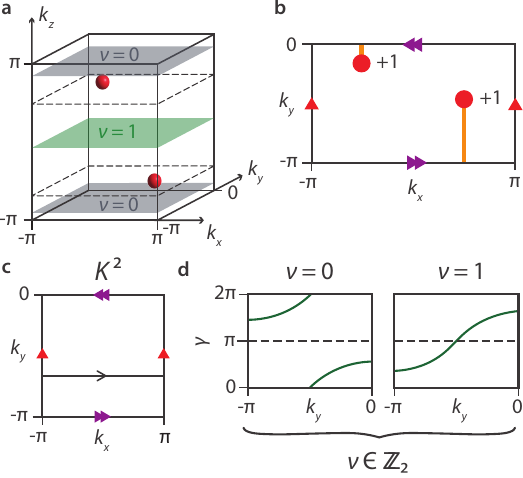}}
    \caption{%
        \textbf{a},~The distribution of Weyl points (red spheres) at $E=0$ in the fundamental domain, $K^2\times S^1$, for the model in \cref{eq:d_functions}. All Weyl points have Chern number $+1$. 
        Highlighted $k_z$ planes indicate that an odd number of Weyl points mediates the transitions between different values of the $\mathbb{Z}_2$ invariant $\nu$, discussed below.
        \textbf{b},~Fermi arcs obtained on truncating the system along the $z$ direction, connecting projections of Weyl points (red circles) with charges of the same sign via an orientation-reversing path. The associated fundamental surface Brillouin zone forms a Klein bottle.
         \textbf{c},~Berry phase, $\gamma$, calculated on the Klein bottle, $K^2$. The Berry connection is integrated along $k_x$ and plotted as a function of $k_y$ from $-\pi$ to $0$. 
        \textbf{d},~The invariant $\nu$ gives a $\mathbb{Z}_2$ classification on $K^2$, the transitions of which are mediated by Weyl points as shown in \textbf{a}.
        }
    \label{fig:figure2}
\end{figure}

Next, we consider the consequences of this symmetry using a two-band, spinless model on the three-dimensional cubic lattice with a Bloch Hamiltonian of the form
\begin{align}
    H(\mathbf{k}) = \mathbf{d}\cdot \boldsymbol{\sigma} = d_x(\mathbf{k}) \sigma_x + d_y(\mathbf{k}) \sigma_y + d_z(\mathbf{k}) \sigma_z,
    \label{eq:Bloch_H}
\end{align}
where $\sigma_{x,y,z}$ are the Pauli matrices, and the components of $\mathbf{d}(\mathbf{k}) = [d_x, d_y, d_z](\mathbf{k})$ are individually subject to \cref{eq:H_cons}.
As a concrete example, we take:
\begin{align}
    \begin{split}
    \label{eq:d_functions}
    d_x(\mathbf{k}) &= \cos{k_x}, \\
    d_y(\mathbf{k}) &= \sin{k_z} - \sin{k_x}\sin{k_y}  - \tfrac{1}{2}, \\
    d_z(\mathbf{k}) &= \cos{k_z} + \sin{k_x}\cos{k_y} + 1
    \end{split}
\end{align}
and note that, physically, the constraints on $\mathbf{d}(\mathbf{k})$ imply a suppression of certain hoppings in real space~\cite{SupplementalMaterial}.

The bands touch at Weyl points where $|\mathbf{d}(\mathbf{k})|=0$. 
For our model, we find two Weyl points in the $K^2\times S^1$ fundamental domain (\cref{fig:figure2}a). 
Their chiralities can be computed by enclosing each Weyl point within a spherical shell and integrating the Berry curvature flux through it~\cite{Vanderbiltbook}.
However, since $K^2\times S^1$ is non-orientable, there is no globally consistent orientation. This is relevant for the calculation of the Chern number since it is a pseudoscalar, and therefore flips sign upon orientation reversal. 
Thus, while an absolute sign for the chirality cannot be established, an orientation choice can be made on any finite region that does not include the orientation-reversing planes $k_y = -\pi$ and $k_y = 0$, such that it contains all the Weyl points. 
This choice can be used to assign the signs of their charges through a Berry curvature flux integral, which provides an unambiguous definition of relative chirality within the region. 
Interestingly, this implies that non-orientability allows all Weyl points on the manifold to have the same relative chirality. 
For example, if we start with two Weyl points of opposite chiralities, we can change the chirality of one of them by moving it along an orientation-reversing path, \ie one that crosses the planes $k_y = -\pi$ and $k_y = 0$ an odd number of times, resulting in both Weyl points having the same relative chirality.
This is similar to the chirality flip observed in non-Hermitian systems when Weyl points encircle exceptional nodal lines~\cite{Sun2020, Wojcik2020}.

In our model, we find that both Weyl points on the $K^2\times S^1$ manifold carry a charge of $+1$ (\cref{fig:figure2}a).
Evidently, the Nielsen--Ninomiya theorem is circumvented on the fundamental domain~\footnote{We note that the total chirality on $T^3$ is vanishing and therefore the Nielsen--Ninomiya theorem holds on $T^3$ even if it is violated on $K^2 \times S^1$.} since the total chirality is $\chi = +2$.
In~\cite{SupplementalMaterial}, we show that this circumvention is possible because $\mathbf{d}(\mathbf{k})$ is discontinuous on $K^2\times S^1$, even though the Hamiltonian itself is continuous.
This discontinuity is directly tied to the non-orientability of the underlying manifold, and renders the Nielsen--Ninomiya theorem inapplicable. 
We also further explore the nature of this circumvention in~\cite{SupplementalMaterial}: first, we show a direct physical consequence of this circumvention---systems that exhibit a non-zero total chirality on $K^2\times S^1$ necessarily host gapless surface states where twisted boundary identifications are made; and second, we prove that fine-tuning $\mathbf{d}(\mathbf{k})$, such that it is continuous on the fundamental domain, restores the Nielsen--Ninomiya theorem.

Physically, the relative chiralities of Weyl points can also be ascertained through the Fermi arcs: Weyl points of the same chirality are connected via Fermi arcs that lie on orientation-reversing paths (\cref{fig:figure2}b); whereas Weyl points with opposite chirality are connected via Fermi arcs that lie on orientation-preserving paths, \ie those that intersect the lines $k_y = -\pi$ and $k_y = 0$ an even number of times. 
In~\cite{SupplementalMaterial}, we show that these features are general by considering a different non-orientable manifold, the real projective plane $RP^2$.

We will now show that Weyl points on non-orientable manifolds carry an additional $\mathbb{Z}_2$ topological charge, which results in a different no-go theorem. 
To identify the $\mathbb{Z}_2$ charge, we consider topological invariants on two-dimensional gapped subspaces of the three-dimensional Brillouin zone. 
Explicitly, we consider fixed-$k_z$ subspaces restricted to the fundamental domain, which form Klein bottles, $K^2$ (\cref{fig:figure1}{b, c}).
We can then integrate the Berry connection along the $k_x$ direction to obtain the Berry phase $\gamma(k_y)$ (\cref{fig:figure2}c).
Since $k_y = -\pi$ and $k_y = 0$ are related by a $k_x$-mirror operation, integrating the Berry connection along these lines leads to a relative minus sign, and therefore $\gamma(k_y = -\pi) = -\gamma(k_y = 0) \pmod{2\pi}$. 
By counting the number of crossings $W_\pi$ of the Berry phase through the horizontal line $\gamma = \pi$, it can be shown that curves with a given $W_\pi$ parity can be deformed into one another but cannot be deformed into those with a different parity.
This defines a $\mathbb{Z}_2$ invariant $\nu \equiv W_\pi \bmod 2 \in \{0, 1\}$ on $K^2$~\cite{Chen2022} (\cref{fig:figure2}d). 
Similar to the Chern number, $\nu$ originates from the topological classification of line bundles over their fundamental domains, which can be expressed in terms of the second cohomology groups $H^2(T^2; \mathbb{Z}) = \mathbb{Z}$ and $H^2(K^2; \mathbb{Z}) = \mathbb{Z}_2$. 
Therefore, the classification is stable under the addition of trivial bands.
Furthermore, this invariant leads to edge states when interfaced with a trivial system~\cite{Chen2022, SupplementalMaterial}.

We now compute the value of $\nu$ for various fixed-$k_z$, Klein-bottle  cuts for the model in \cref{eq:d_functions}. 
\cref{fig:figure2}a shows that the value of $\nu(k_z)$ changes by unity as $k_z$ passes through an odd number of Weyl points. 
This suggests that a non-trivial value of $\nu$ is associated with Weyl points of odd chirality. 
We show in~\cite{SupplementalMaterial} that a local Berry phase calculation for the $\mathbb{Z}_2$ invariant can be carried out by enclosing the Weyl point within a \emph{two-sided} Klein bottle~\cite{woll1971one}.
This shows that $\nu$ is indeed sourced by Weyl points and, accordingly, we may associate a $\mathbb{Z}_2$ charge to each. 
The same conclusion can be reached by relating $\nu$ to the Chern number $C$ of the Weyl point by noticing that the Berry phase jumps by $2\pi C$ at the momentum of the Weyl point~\cite{Yoshida.1:2023, Yoshida.2:2023}. This leads to the relation $\nu = C \bmod 2$~\cite{SupplementalMaterial}. 

\begin{figure*}
    \centerline{%
    \includegraphics[scale=1]{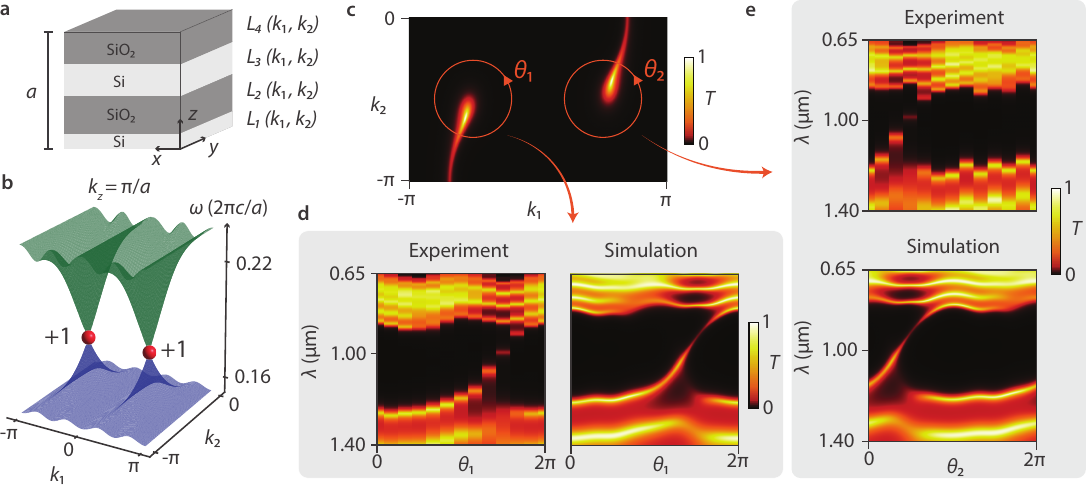}}
    \caption{%
        \textbf{a},~The unit cell of the PhCs that consist of four layers of thicknesses $L_1$ to $L_4$. The lattice constant in the $z$-direction is $a = \SI{200}{nm}$. 
        \textbf{b},~The lowest two bands of the PhCs at $k_z = \pi/a$, as a function of $k_1$ and $k_2$. The two Weyl points, shown as red spheres, both carry a Chern number of $+1$. These are ``ideal'' Weyl points that are frequency-isolated from the other bands. 
        \textbf{c},~Fermi arcs on the top surface of the finite-in-$z$ system on the $K^2$ surface Brillouin zone formed by the parameters $k_1$ and $k_2$. 
        (\textbf{d},\textbf{e})~Experimentally obtained and simulated transmission spectra showing the Fermi-arc surface state dispersion along circular loops enclosing the Weyl points. 
        The loops around the projections of the Weyl points are parametrized by angular variables $\theta_1$ and $\theta_2$. 
        }
    \label{fig:figure4}
\end{figure*}

Having uncovered that Weyl points carry a $\mathbb{Z}_2$ charge, we now show that this charge is subject to a no-go theorem on $K^2\times S^1$. 
The argument proceeds as follows: since the $k_z$ direction is periodic, $\nu(k_z = -\pi) = \nu(k_z = \pi)$. 
This forbids the presence of a net non-zero $\mathbb{Z}_2$ charge of the Weyl points within $K^2\times S^1$. 
To show this, we consider the case of a single Weyl point at the $k_z = 0$ plane. 
Due to this Weyl point, the value of $\nu(k_z = 0^-)$ would differ from $\nu(k_z = 0^+)$ by unity. 
If there are no other Weyl points, it follows that the values, $\nu(k_z = -\pi) = \nu(k_z = 0^-)$ and $\nu(k_z = \pi) = \nu(k_z = 0^+)$ would not be equal, which is not possible if the $k_z$-direction is periodic. 
Thus, a second Weyl point is needed to ensure that the invariants at the two $k_z$ planes match. 
Accordingly, the total $\mathbb{Z}_2$ charge of the Weyl points on $K^2\times S^1$ must vanish; or, equivalently, the sum of Chern numbers of the Weyl points must be even---but as we have shown, not necessarily vanishing. 
This no-go theorem implies that for systems obeying~\cref{eq:H_cons} the minimum number of singly-charged Weyl points is two on $K^2\times S^1,$ and four on the toroidal Brillouin zone, even under broken time-reversal symmetry. 
In~\cite{SupplementalMaterial}, we further discuss the roles of inversion and time-reversal symmetries, and provide a second argument for this no-go theorem based on the configuration of Fermi arcs. 

In this final section, we turn our attention to an experimental demonstration of Weyl points on non-orientable manifolds, which we realize in a photonic system endowed with synthetic momenta. 
The system considered here consists of a family of optical multilayer structures, \ie one-dimensional photonic crystals (PhCs). 
Their unit cells are composed of four dielectric layers alternating between two materials, Silicon $(\varepsilon_{\mathrm{Si}} = 12.5)$ and SiO$_2$ $(\varepsilon_{\mathrm{SiO}_2} = 2.25)$, with a lattice constant $a$ in the $z$-direction (\cref{fig:figure4}a). 
Light propagation in such PhCs is governed by a Maxwell eigenvalue problem for the electromagnetic field eigenmodes, and their corresponding frequency eigenvalues, analogous to electrons moving in a crystalline solid~\cite{Joannopoulos2008, Sakoda_PhC}.  
When light propagation along the $z$-direction is considered, this is reduced to a one-dimensional problem which is solved using Bloch's theorem. 
The resulting field solutions form discrete frequency bands as a function of the quasimomentum, $k_z$, which may be separated by photonic band gaps. 

We introduce two periodic parameters, $k_1$ and $k_2$, to modulate the thicknesses, $L_1$, $L_2$, $L_3$ and $L_4$, of each of the four layers in the unit cell. 
The parameters $k_1, k_2$ serve as synthetic momentum degrees of freedom which, along with the quasimomentum $k_z$, result in a three-dimensional toroidal parameter space within which Weyl points can exist~\cite{Wang2017, nguyen2023fermi}. 
We choose $L_1$ to $L_4$ such that the non-symmorphic symmetry given in \cref{eq:H_cons} is satisfied in $(k_1,k_2,k_z)$-space. 
In particular, we choose $L_1 = \tfrac{1}{4}a (1+\cos k_1), L_2 = \tfrac{1}{4}a (1+\sin k_1 \cos k_2), L_3 = \tfrac{1}{4}a (1-\cos k_1), L_4 = \tfrac{1}{4}a (1-\sin k_1 \cos k_2)$.
Thus the fundamental domain is $K^2\times S^1$ after making boundary identifications at the $k_2 = -\pi$ and $0$ planes.

We find that the fundamental domain hosts two Weyl points between the lowest two bands of this system (\cref{fig:figure4}b), each with a relative chirality of $+1$. 
This implies that they each carry a $\mathbb{Z}_2$ charge of $\nu = 1$, as we explicitly show in \cite{SupplementalMaterial}.
The total chirality of the Weyl points therefore does not vanish, similar to what was observed in the tight-binding model in \cref{eq:d_functions}.
However, the total $\mathbb{Z}_2$ charge vanishes, consistent with the no-go theorem for these charges. 
The higher bands can host increasingly larger numbers of Weyl points, while always maintaining a vanishing total $\mathbb{Z}_2$ charge. 
We discuss the many-band case in more detail in~\cite{SupplementalMaterial}.

On truncating the PhCs along the $z$-direction, Fermi arcs are expected to emerge from the projections of the Weyl points in the surface Brillouin zone formed by $(k_1, k_2)$. 
Since the Fermi arcs are localized on the surfaces, they possess an enormous linewidth generated by the strong out-coupling to plane waves in the air above the PhCs. 
To remedy this, we clad the PhCs with additional layers on the top surface to better confine these states~\cite{SupplementalMaterial}. 
Doing so allows for the observation of the Fermi arcs in the transmission spectrum of the PhCs, a simulation of which is shown in \cref{fig:figure4}c. 
When the dispersion of the Fermi arcs is plotted along a loop that encloses the projection of a Weyl point, these states fully cross the band gap, with the direction of their spectral flow determined by the sign of the chirality of the enclosed Weyl point. Since our Weyl points have the same chirality, we expect the same spectral flow pattern for both nodes as simulated in \cref{fig:figure4}{d, e}.

For the experiment, we fabricate a series of PhCs that correspond to values of $k_1$ and $k_2$ lying on the loops that enclose the Weyl point projections, as shown in~\cref{fig:figure4}c (further details on the simulations and experiment are given in~\cite{SupplementalMaterial}).
\Cref{fig:figure4}{d, e} show the experimental results along with corresponding simulations. 
We see that the surface states cross the gap with identical spectral flow for both Weyl points, indicating that they carry the same chirality.

In summary, by implementing Weyl quasiparticles in lattice models with non-symmorphic momentum-space symmetries, we have explored their fate on non-orientable manifolds. 
On the associated fundamental domain, the Hamiltonian, its eigenstates, and all physical observables are continuous. 
However, we have shown that the chirality of Weyl points need not sum to zero, circumventing the Nielsen--Ninomiya theorem. 
The underlying non-orientable domain endows the Weyl points with an additional $\mathbb{Z}_2$ charge, whose conservation enforces a new no-go theorem. 
Finally, we have experimentally demonstrated the phenomenology of such Weyl points in a photonic platform with synthetic momenta.
Our work suggests several new research directions. 
For example, one can consider other non-orientable manifolds in dimensions two and higher that might host their own unique topological invariants and new gapless points~\cite{tao2023higherorder, jinbing2023synthetic, li2023klein}. It will also be interesting to explore the properties of Landau levels originating from both real and pseudo-magnetic fields when the chirality does not vanish~\cite{ilan2020pseudo, behrends2019landau, barsukova2024direct}. 
More broadly, this opens up new avenues to explore how other chiral objects, such as multifold fermions~\cite{Manes2012, Bradlyn2016} and exceptional points~\cite{Bergholtz2021, Konig2023}, fare in non-orientable settings. We believe that the approaches introduced here may help answer these fundamental questions.

\vskip 1.5ex
We thank Clifford Taubes, Terry A.~Loring, Adolfo G.~Grushin, and Jonathan Guglielmon for stimulating discussions. 
A.G.F.\ acknowledges support from the Henry W. Kendall Fellowship and the Whiteman Fellowship, and thanks the University of S\~ao Paulo for its hospitality, where part of this work was completed. 
T.C.\ acknowledges the support of a research grant (project no.~42106) from Villum Fonden. 
S.V., M.C.R., T.L.H., and M.S.\ acknowledge support from the U.S.\ Office of Naval Research (ONR) Multidisciplinary University Research Initiative (MURI) under Grant No.\ N00014-20-1-2325 on Robust Photonic Materials with Higher-Order Topological Protection.
This material is based upon work also supported in part by the U. S. Army Research Office through the Institute for Soldier Nanotechnologies at MIT, under Collaborative Agreement Number W911NF-23-2-0121.
This work was carried out in part through the use of MIT.nano's facilities.
\vskip 1 in

\bibliography{bibliography}

\end{document}


\title{\texorpdfstring{
        SUPPLEMENTAL MATERIAL\\[1ex]
         Weyl points on non-orientable manifolds
        }
        {Supplemental Material}
       }

\author{Andr\'e Grossi Fonseca}
\email{agfons@mit.edu}
\affiliation{\mitaffil}

\author{Sachin Vaidya}
\email{svaidya1@mit.edu}
\affiliation{\mitaffil}

\author{Thomas Christensen}
\affiliation{\dtuaffil}

\author{Mikael C.~Rechtsman}
\affiliation{\pennaffil}

\author{Taylor L.~Hughes}
\affiliation{\uiucaffil}

\author{Marin Solja\v{c}i\'{c}}
\affiliation{\mitaffil}

\maketitle

\setlength{\parindent}{0em}
\setlength{\parskip}{.5em}

\noindent{\small\textbf{\textsf{CONTENTS}}}\\ 
\twocolumngrid
\begingroup
    \let\bfseries\relax 
    \deactivateaddvspace 
    \deactivatetocsubsections 
    \makeatletter\@starttoc{toc}\makeatother 
\endgroup
\onecolumngrid

\count\footins = 1000 
\interfootnotelinepenalty=10000 

\section{Glide symmetry from orbital displacement}

In the main text, we motivated our discussion of the momentum-space glide symmetry in analogy to translational symmetry. 
We imposed a unitary-free symmetry because the Bloch Hamiltonian is, in a suitable basis, periodic on the reciprocal lattice, with no unitary conjugation.
However, this strict periodicity is only valid if the unit-cell degrees of freedom have no spatial displacement, or if one considers the so-called ``periodic Bloch Hamiltonian''~\cite{Cayssol2021}.
In general, the Bloch Hamiltonian is only periodic up to a unitary, if one adopts the ``canonical Bloch Hamiltonian''~\cite{Cayssol2021}, also known as the ``periodic gauge condition''~\cite{Vanderbiltbook}.
Below, we show that if one applies this formalism to the glide symmetry, there is also a unitary that appears for the canonical Bloch Hamiltonian, but all general statements that are subsequently made on Weyl points remain true. 
Furthermore, we note that working in the convention of the canonical Bloch Hamiltonian leads to greater flexibility in constructing tight-binding models, which could be important for material realizations and for extending our work to other platforms. 
To illustrate the ideas concretely, we also provide a specific example of a simple tight-binding model where the orbital degrees of freedom are displaced from the origin of each unit cell. 

Assuming periodic translation symmetry, one can derive the unitary relation by starting from the periodic Bloch Hamiltonian:
\begin{align}
    H_{\mathrm{p}}(\mathbf{k}) = \e{-\iu \mathbf{k} \cdot \mathbf{R}} H \e{\iu \mathbf{k} \cdot \mathbf{R}}, 
    \label{eq:periodic_bloch}
\end{align}
where $\mathbf{R}$ is the position operator associated with each unit cell center and $H$ is the real-space Hamiltonian with the periodicity of the Bravais lattice.
By shifting the crystal momentum by a reciprocal lattice vector $\mathbf{G}$, 
\begin{align}
    H_{\mathrm{p}}(\mathbf{k} + \mathbf{G}) = \e{-\iu \mathbf{G} \cdot \mathbf{R}} H_{\mathrm{p}}(\mathbf{k}) \e{\iu \mathbf{G} \cdot \mathbf{R}}. 
\end{align}
Then, demanding periodicity with the reciprocal space $H_{\mathrm{p}}(\mathbf{k} + \mathbf{G}) = H_{\mathrm{p}}(\mathbf{k})$ yields the familiar condition
\begin{align}
    \e{-\iu \mathbf{G} \cdot \mathbf{R}} = 1,
    \label{eq:trans_condition}
\end{align}
which defines the allowed reciprocal lattice vectors.
We can then employ this condition to find the unitary condition on the canonical Bloch Hamiltonian, defined as:
\begin{align}
    H_{\mathrm{c}}(\mathbf{k}) = \e{-\iu \mathbf{k} \cdot \mathbf{r}} H \e{\iu \mathbf{k} \cdot \mathbf{r}}, 
    \label{eq:canonical_bloch}
\end{align}
where now $\mathbf{r}$ is the full position operator.
Shifting momentum once again, we get
\begin{align}
    H_{\mathrm{c}}(\mathbf{k} + \mathbf{G}) = \e{-\iu \mathbf{G} \cdot \mathbf{r}} H_{\mathrm{c}}(\mathbf{k}) \e{\iu \mathbf{G} \cdot \mathbf{r}}. 
\end{align}
Then, writing $\mathbf{r} = \mathbf{R} + \boldsymbol{\delta}$, with $\boldsymbol{\delta}$ being the position operator within the unit cell, and employing \cref{eq:trans_condition}, we arrive at
\begin{align}
    H_{\mathrm{c}}(\mathbf{k} + \mathbf{G}) = U^\dag H_{\mathrm{c}}(\mathbf{k}) U, 
    \quad \text{with } U = \e{\iu \mathbf{G} \cdot \boldsymbol{\delta}},
\end{align}
\ie the canonical Bloch Hamiltonian is not manifestly periodic on the reciprocal lattice, unlike the periodic Bloch Hamiltonian.

Next, we would like to ascertain the effect of an orbital displacement in the unit cell, \ie a non-zero $\boldsymbol{\delta}$, on the momentum-space glide symmetry studied in the main text.
We shall follow a similar procedure to the one above, first by imposing the glide symmetry exactly on the periodic Bloch Hamiltonian, and then studying the consequences for the corresponding canonical Bloch Hamiltonian.
We then start with the assumption that
\begin{align}
    H_{\mathrm{p}}(k_x, k_y) = H_{\mathrm{p}}(-k_x, k_y+\pi).
\end{align}
Writing both sides in real space, this condition amounts to
\begin{equation}
    \e{-\iu (k_x R_x + k_y R_y)} H \e{\iu (k_x R_x + k_y R_y)} = 
    \e{-\iu (-k_x R_x + (k_y+\pi) R_y)} H \e{\iu (-k_x R_x + (k_y+\pi) R_y)}.
    \label{eq:glide_condition}
\end{equation}
Under this symmetry, the canonical Bloch Hamiltonian can be written
\begin{align}
    \begin{split}
        H_{\mathrm{c}}(-k_x, k_y+\pi) &= \e{-\iu (-k_x (R_x+ \delta_x) + (k_y+\pi) (R_y+\delta_y))} 
        H \e{\iu (-k_x (R_x+ \delta_x) + (k_y+\pi) (R_y+\delta_y))} \\
        &= \e{-\iu (-k_x \delta_x + (k_y + \pi)\delta_y)}\e{-\iu (-k_x R_x + (k_y+\pi) R_y)} 
        H \e{\iu (-k_x R_x + (k_y+\pi)R_y)} \e{\iu (-k_x \delta_x + (k_y + \pi)\delta_y)}.
    \end{split}
\end{align}
Finally, we can use \cref{eq:glide_condition} to simplify this to
\begin{align}
    \begin{split}
        H_{\mathrm{c}}(-k_x, k_y+\pi) 
        &= \e{-\iu (-k_x \delta_x + (k_y + \pi)\delta_y)}\e{-\iu (k_x R_x + k_y R_y)} 
        H \e{\iu (k_x R_x + k_y R_y)} \e{\iu (-k_x \delta_x + (k_y + \pi)\delta_y)} \\
        &= \e{\iu(2k_x \delta_x-\pi \delta_y)} \e{-\iu \mathbf{k} \cdot \mathbf{r}} H \e{\iu \mathbf{k} \cdot \mathbf{r}} \e{-\iu(2k_x \delta_x-\pi \delta_y)} \\
        &= \Tilde{U}^\dag H_{\mathrm{c}}(k_x, k_y) \Tilde{U},
    \end{split}
    \label{eq:H_canonical_cons}
\end{align}
where $\Tilde{U} = \e{\iu\Delta\mathbf{k} \cdot \boldsymbol{\delta}}$ is a unitary operator, dependent on the ``relative'' momentum $\Delta\mathbf{k} = (-k_x, k_y +\pi) - (k_x, k_y)$ and the intra-cell position operator $\boldsymbol{\delta}=(\delta_x, \delta_y)$.
Therefore, there is a unitary relationship here as well, which simplifies to the unitary-free symmetry we studied if $\boldsymbol{\delta} = \mathbf{0}$, in precise analogy to translational symmetry. 

\begin{figure}
    \centerline{%
    \includegraphics[scale=1]{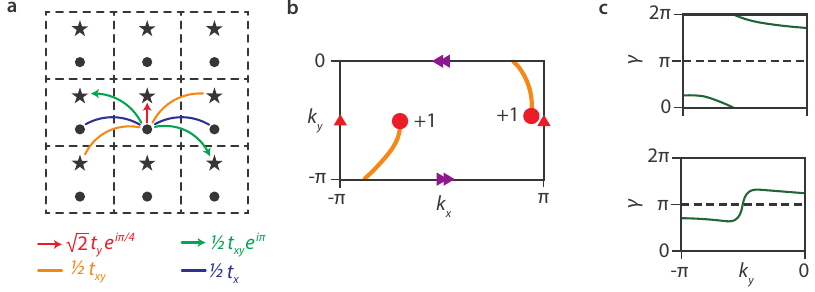}}
    \caption{%
        \textbf{a}, In-plane hopping profile for the model in~\cref{eq:hoppings}. Hopping strengths and phases are shown below, in terms of the model parameters.
        \textbf{b},~Fermi arcs obtained on truncating the system along the $z$ direction, connecting projections of Weyl points with charges of the same sign via an orientation-reversing path.
        \textbf{c},~Berry phase, $\gamma$, calculated on $K^2$ for $k_z$ cuts below (upper) and above (lower) a Weyl point, indicating that $\nu(k_z)$ shifts by unity upon sweeping across a Weyl point.
        }
    \label{fig:figure9}
\end{figure}

We now illustrate this discussion with an explicit model. 
We consider a square lattice with two orbitals located at $(0, \pm 1/4)$, as shown in \cref{fig:figure9}a.
In this case, $\Tilde{U} = \e{\iu \pi \delta_y} = \text{diag}(1, \iu)$.
Writing the canonical Bloch Hamiltonian as $H_{\mathrm{c}}(\mathbf{k}) = \mathbf{d}(\mathbf{k})\cdot \boldsymbol{\sigma}$, the symmetry in \cref{eq:H_canonical_cons} requires that
\begin{align}
    \begin{split}
        &d_x(-k_x, k_y + \pi, k_z) = d_y(k_x, k_y, k_z), \\
        &d_y(-k_x, k_y + \pi, k_z) = -d_x(k_x, k_y, k_z), \\
        &d_z(-k_x, k_y + \pi, k_z) = d_z(k_x, k_y, k_z).
    \end{split}
\end{align}
As a concrete example, we take:
\begin{align}
    \begin{split}
        d_x(\mathbf{k}) &= t_y (\cos \tfrac{1}{2}k_y - \sin \tfrac{1}{2}k_y) + t_{xy}\cos(k_x+\tfrac{1}{2}k_y) + t_{yz}\sin k_z \sin \tfrac{1}{2}k_y , \\
        d_y(\mathbf{k}) &= -t_y ( \sin \tfrac{1}{2}k_y + \cos \tfrac{1}{2}k_y) + t_{xy} \sin(k_x-\tfrac{1}{2}k_y) + t_{yz}\sin k_z \cos \tfrac{1}{2}k_y, \\
        d_z(\mathbf{k}) &= t_x \cos{k_x} + t_z \cos{k_z}.  
    \end{split}
    \label{eq:hoppings}
\end{align}

The in-plane hopping strengths and phases in real space are shown in \cref{fig:figure9}a.
The out-of-plane hoppings are between nearest-neighbors and next-nearest-neighbors, and are parameterized by $t_z$ and $t_{yz}$, respectively.
By choosing \eg $t_y = t_z = t_{yz} = 1, t_x = 0.5, t_{xy} = 1.5$, this model has two Weyl points on the $K^2 \times S^1$ fundamental domain, both with chirality $+1$, which are connected by a Fermi arc lying on an orientation-reversing path (\cref{fig:figure9}b).
Finally, we also plot the Berry phase on different $k_z$ cuts below and above the Weyl points (\cref{fig:figure9}c), showing that the $\mathbb{Z}_2$ invariant $\nu(k_z)$ changes by unity as $k_z$ passes through an odd number of Weyl points.
Therefore, this model illustrates that the physics discussed in the main text remains intact if we allow for the more general unitary symmetry in \cref{eq:H_canonical_cons}, as is the case with translational symmetry. 

\section[Terms compatible with $\text{\emph{K}}^{\text{2}}\times \text{\emph{S}}^{\text{1}}$ fundamental domain]{General terms compatible with a $\text{\emph{K}}^{\text{2}}\times \text{\emph{S}}^{\text{1}}$ fundamental domain}

As we have shown in the main text, the momentum-space fundamental domain of a three-dimensional lattice is reduced from the torus $T^3$ to $K^2 \times S^1$ if the Bloch Hamiltonian obeys the following constraint:
\begin{align}
    H(k_x, k_y, k_z) = H(-k_x, k_y+ \pi, k_z).
    \label{eq:H_cons}
\end{align}
This constraint restricts the in-plane sinusoidal terms that can appear in the Hamiltonian to be one of those in \cref{tab:allowed terms}. 
For example, couplings to the neighboring unit cells along $y$, which in momentum space would give rise to terms such as $\cos(k_y)$ or $\sin(k_y)$, are disallowed by this symmetry. 
This is consistent with the physical expectation that a reduction of the momentum-space fundamental domain---in our case, from $T^3$ to $K^2 \times S^1$---would lead to an extended range of couplings in real space.

\begin{table}[tb]
    \begin{tabular}{c c c} 
     \toprule
     \multirow{2}{*}{Term} & \multicolumn{2}{c}{Parity}    \\
      & $n$ & $m$ \\ 
     \midrule
     $\cos nk_x$ & any &  \\
     $\cos nk_y$ & even &  \\
     $\sin nk_y$ & even &    \\
     $\sin nk_x\sin mk_y$ & any & odd  \\
     $\sin nk_x\cos mk_y$ & any & odd  \\
     \bottomrule
    \end{tabular}
    \caption{Allowed coupling terms in the Bloch Hamiltonian subject to the constraint of \cref{eq:H_cons}.}
    \label{tab:allowed terms}
\end{table}

The symmetry in \cref{eq:H_cons} does not involve unitary conjugation, unlike, e.g., spatial symmetries. This implies that the $\sigma$ matrices in $H(\mathbf{k}) = \mathbf{d}(\mathbf{k})\cdot \boldsymbol{\sigma}$ do not transform under this symmetry, and therefore the terms in \cref{tab:allowed terms} can enter any component of $\mathbf{d}(\mathbf{k})$.
This remains true for $N$-band models, where now $\boldsymbol{\sigma}$ is to be understood as a set of generalized Gell-Mann matrices.

\section{Weyl points on $\text{\emph{RP}\textsuperscript{2}}\times \text{\emph{S}\textsuperscript{1}}$}

\begin{figure*}[ht]
    \centerline{%
    \includegraphics[scale=1]{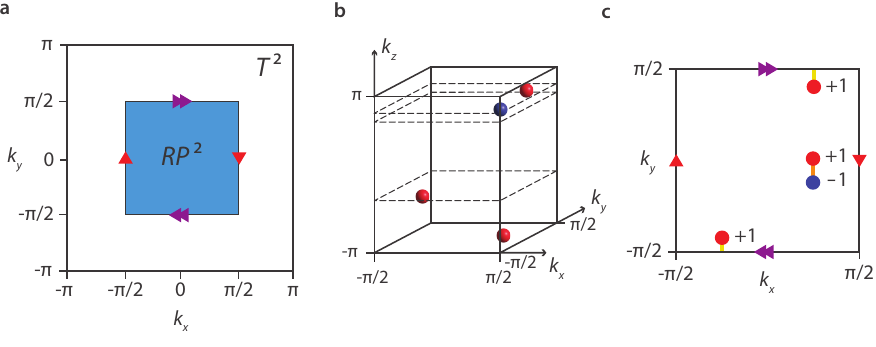}}
    \caption{
        \textbf{a},~In the case of Hamiltonians obeying the symmetry in \cref{eq:H_cons_RP2}, the fundamental domain in momentum space takes the form of a non-orientable manifold, $RP^2\times S^1$. 
        Two-dimensional cuts in the $(k_x,k_y)$ plane form a real projective plane, $RP^2$.
        Arrows indicate boundary identifications.
        \textbf{b},~The distribution of Weyl points (red and blue spheres) at $E=0$ in the fundamental domain for the model in \cref{eq:d_functions_RP2}.
        Weyl points in red/blue have Chern number $+1$/$-1$.
        \textbf{c},~Fermi arcs obtained on truncating the system along the $z$ direction, connecting projections of Weyl points. Weyl points of the same chirality are connected via orientation-reversing paths and those of opposite chirality are connected via orientation-preserving paths.
        }
    \label{fig:figure1}
\end{figure*}

Our construction of the $K^2 \times S^1$ fundamental domain relied on the momentum-space nonsymmorphic symmetry defined in \cref{eq:H_cons}, consisting of a reflection about $k_x$ and a $\pi$-shift in $k_y$. 
If we impose an additional symmetry, where the roles of $k_x$ and $k_y$ are swapped, \ie
\begin{align}
    H(k_x, k_y, k_z) = H(-k_x, k_y+ \pi, k_z) = H(k_x+ \pi, -k_y, k_z),
    \label{eq:H_cons_RP2}
\end{align}
the resulting fundamental domain forms a different non-orientable manifold. This manifold is obtained by gluing and twisting two pairs of planes (\cref{fig:figure1}a). The torus $T^3$ is then subdivided into four copies of the manifold $RP^2 \times S^1$, where $RP^2$ is the real projective plane.

As in the main text, we consider two-band Bloch Hamiltonians of the form $H(\mathbf{k}) = \mathbf{d}(\mathbf{k})\cdot \boldsymbol{\sigma}$, now subject to the symmetry in \cref{eq:H_cons_RP2}.
Concretely, we take:
\begin{align}
    \begin{split}
    \label{eq:d_functions_RP2}
    d_x(\mathbf{k}) &= \cos{2k_x}, \\
    d_y(\mathbf{k}) &= \cos{2k_y}\sin{k_z} + \sin{k_x}\sin{k_y} - \cos{k_z} - 1, \\
    d_z(\mathbf{k}) &= \cos{k_z} + \sin{2k_x}\cos{k_y} + \sin{2k_y}\cos{k_x}.
    \end{split}
\end{align}
This model has four Weyl points on the $RP^2 \times S^1$ manifold, three of which have charge $+1$ and one with charge $-1$ (\cref{fig:figure1}b). 
The total chirality in the fundamental domain is therefore $\chi = +2$, and the Nielsen--Ninomiya theorem is also circumvented here.
Note that the Weyl point with charge $-1$ connects to one with charge $+1$ through a Fermi arc lying on an orientation-\emph{preserving} path, whereas the two other Weyl points connect via orientation-reversing Fermi arcs, in analogy to the model in the main text on $K^2 \times S^1$ (\cref{fig:figure1}c).

\begin{figure*}
    \centerline{%
    \includegraphics[scale=1]{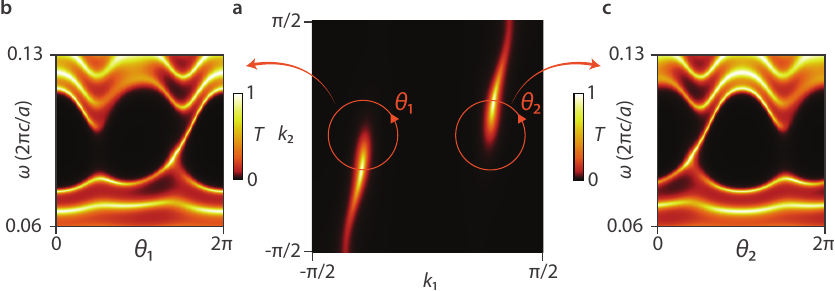}}
    \caption{
        \textbf{a},~Fermi arcs on the top surface of the finite-in-$z$ system on the $RP^2$ surface Brillouin zone formed in $(k_1, k_2)$ space with layer thicknesses modulated as in \cref{eq:L_functions_RP2}.
        (\textbf{b},\textbf{c})~Simulated transmission spectra showing the Fermi-arc surface state dispersion along circular loops enclosing the Weyl points. 
        The loops around the projections of the Weyl points are parametrized by angular variables $\theta_1$ and $\theta_2$. Both Weyl points have the same relative chirality, as indicated by the similar spectral flow of the Fermi arcs.
        }
    \label{fig:figure2}
\end{figure*}

We can also investigate the physics of Weyl points on the $RP^2 \times S^1$ manifold using photonic crystals with synthetic momenta. To do this, we choose functions modulating layer thicknesses that are invariant under the symmetry in \cref{eq:H_cons_RP2}. In particular, we choose
\begin{align}
    \label{eq:L_functions_RP2}
    \begin{split}
    L_1(k_1, k_2) &= \tfrac{1}{2}a (1+\cos 2k_1), \\
    L_2(k_1, k_2) &= \tfrac{1}{2}a (1+\sin k_1 \sin k_2), \\
    L_3(k_1, k_2) &= \tfrac{1}{2}a (1-\cos 2k_1), \\
    L_4(k_1, k_2) &= \tfrac{1}{2}a (1-\sin k_1 \sin k_2).
    \end{split}
\end{align}
This system hosts two Weyl points with charge $+1$ between the two lowest photonic bands.
Their identical relative chirality can be seen by either noting that they are connected by a Fermi arc lying on an orientation-reversing path (\cref{fig:figure2}a), or by calculating the spectral flow of the Fermi arcs along closed loops enclosing the Weyl points (\cref{fig:figure2}{b, c}).  

\section{Continuity of \MakeLowercase{\textbf{d}(\textbf{k})} on $\text{\emph{K}}^{\text{2}}\times \text{\emph{S}}^{\text{1}}$ and Nielsen--Ninomiya}

We take a deeper look at the Nielsen--Ninomiya theorem to understand why it is circumvented in our case. For manifolds, a simple proof of the Nielsen--Ninomiya theorem can be formulated in terms of the Poincar\'e--Hopf theorem from differential topology~\cite{Karsten1981}. 
This theorem states that for a continuous vector field, the global sum of the indices of all its isolated zeros equals the Euler characteristic of the underlying manifold on which the vector field is defined. 
Here, the vector field is $\mathbf{d}(\mathbf{k})$, the zeros are Weyl points, and the indices characterize the winding of the vector field in the vicinity of the zeros, which is identical to the chiralities of the Weyl points. 
The Euler characteristic of $T^3$ is vanishing, and hence the chiralities of all the Weyl points in the Brillouin zone sum to zero, which concludes the proof for the Nielsen--Ninomiya theorem. 
However, the Euler characteristic of every closed odd-dimensional manifold is also zero, a result that follows from Poincar\'{e} duality~\cite{hatcher2005algebraic}. 
This suggests that the Nielsen--Ninomiya theorem should hold on $K^2\times S^1$, contradicting what we observe for our model in the main text. 

\begin{figure}
    \centerline{%
    \includegraphics[scale=1]{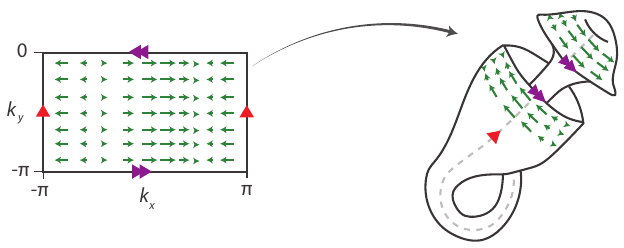}}
    \caption{%
        $d_x(\mathbf{k})\hat{k}_x$ at $k_z = 0$ for the model in the main text. 
        Once the boundary identifications are made along the direction indicated by the purple arrows, $d_x(\mathbf{k})\hat{k}_x$ becomes discontinuous. 
        }
    \label{fig:figure_d_vec}
\end{figure}

The apparent inconsistency can be resolved by noting that, although the Hamiltonian, and hence all physical observables, are continuous across the boundary identifications that generate $K^2\times S^1$, the vector field $\mathbf{d}(\mathbf{k})$ is in fact discontinuous, and thus the Poincar\'e--Hopf and Nielsen--Ninomiya theorems do not apply. 
This can be seen by carefully analyzing the transformation properties of $\mathbf{d}(\mathbf{k}) = d_x(\mathbf{k})\hat{k}_x + d_y(\mathbf{k})\hat{k}_y + d_z(\mathbf{k})\hat{k}_z$ under the momentum-space glide symmetry.
From the $k_x$-mirror operation, $\hat{k}_x$ flips sign, whereas $\hat{k}_y$ and $\hat{k}_z$ are invariant. 
Therefore, continuity of $\mathbf{d}(\mathbf{k})$ at $k_y = -\pi$ and $k_y = 0$ would require $d_x(\mathbf{k})$ to be odd (and $d_{y,z}(\mathbf{k})$ to be even) under the symmetry.
However, since the Hamiltonian is continuous, all components of $\mathbf{d}(\mathbf{k})$, particularly $d_x(\mathbf{k})$, are even under the symmetry, rendering $\mathbf{d}(\mathbf{k})$ discontinuous on $K^2 \times S^1$.
We can visualize this by plotting $d_x(\mathbf{k})\hat{k}_x$ on the fundamental domain and making the boundary identifications (\cref{fig:figure_d_vec}).

The $k_x$-mirror operation is therefore responsible for the discontinuity of $\mathbf{d}(\mathbf{k})$, and for the non-orientability of $K^2\times S^1$.
This implies that the glide symmetry we have considered offers a \emph{generic} mechanism for generating vector field discontinuities on the fundamental domain, resulting in the circumvention of the Nielsen--Ninomiya theorem.

From this discussion, it can be seen that it is possible to construct fine-tuned models in which we require that $d_x(\mathbf{k})$ vanishes along the identified boundaries.
In this case, there is no discontinuity and the Poincar\'e--Hopf theorem is applicable, impliying a vanishing net chirality on $K^2\times S^1$.
Note that this fixes the $k_y$ boundaries of the Klein bottle, which in general can be vertically shifted by any amount (see Section S7). 
To illustrate this, we consider the following example given by 
\begin{align}
    \begin{split}
    \label{eq:d_functions}
    d_x(\mathbf{k}) &= \sin{k_x}\sin{k_y}, \\
    d_y(\mathbf{k}) &= \cos{2 k_y} + \cos{k_x} + 1, \\
    d_z(\mathbf{k}) &= \cos{k_z} + \sin{k_x}\cos{k_y}.
    \end{split}
\end{align}
Here, Weyl points are located at $(k_x, k_y, k_z) = (\pi, -\pi/4, \pm \pi/2)$ and $(\pi, -3\pi/4, \pm \pi/2)$ with charges $\mp 1$ and $\pm 1$, respectively, and hence the Nielsen--Ninomiya theorem holds. We attribute this to the continuity of $\textbf{d}(\textbf{k})$.

We now prove that this is a general feature, \ie that a continuous $\textbf{d}(\textbf{k})$ \emph{guarantees} a vanishing total chirality in the fundamental domain.
First, note that the glide symmetry requires that the Chern numbers on the planes $k_y = -\pi$ and $k_y = 0$ be of the same magnitude but of opposite sign, possibly vanishing.
As we show in Section S5, the statement that the Nielsen--Ninomiya theorem is circumvented on $K^2\times S^1$ is equivalent to these planes being Chern insulators. 
The Berry connection for a two-band model on these planes can be written as~\cite{Bernevig2013}:
\begin{align}
    \mathbf{A}^{\pm}(\mathbf{k}) = \frac{d_x \nabla_\mathbf{k}d_y - d_y \nabla_\mathbf{k}d_x}{2|\mathbf{d}|( |\mathbf{d}| \mp d_z )},  
\end{align}
where $\pm$ denotes the upper/lower band. If $d_x(\mathbf{k})$ is zero, $\mathbf{A}^{\pm}(\mathbf{k})$ vanishes identically and the Chern number is zero on $k_y = -\pi$ and $k_y = 0$. Therefore, requiring a vanishing $d_x(\mathbf{k})$ on these planes implies a vanishing total chirality on $K^2\times S^1$. 

\section[Nielsen--Ninomiya circumvention and surface states]{Nielsen--Ninomiya circumvention and gapless surface states}

\begin{figure*}[ht]
    \centerline{%
    \includegraphics[scale=1]{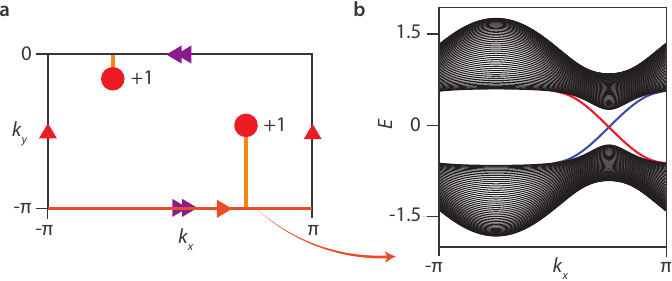}}
    \caption{%
        \textbf{a},~Fermi arcs obtained on truncating the system along the $z$ direction for the model in the main text. These Fermi arcs connect Weyl points with the same relative chirality via an orientation-reversing path.
        \textbf{b},~The dispersion of surface states for $k_y = -\pi$. Due to a non-zero total chirality in the fundamental domain, the $k_y = -\pi$ (and also $k_y = 0$) lines host gapless chiral surface states.
        }
    \label{fig:figure2.1}
\end{figure*}

Here, we discuss a physical consequence of a non-zero total chirality in the fundamental domain $K^2\times S^1$.
In Weyl semimetals, the momentum-space paths that are guaranteed to host chiral states on each surface are those that enclose projections of Weyl points with a non-vanishing total charge.
In general, non-contractible paths in the surface Brillouin zone do not host gapless surface states.
In the case of systems with a non-zero total chirality on $K^2\times S^1$, there are at least two Weyl points of the same relative chirality and they are connected via a Fermi arc lying along an orientation-reversing path.
This path necessarily passes through $k_y = -\pi$ and $k_y = 0$ and thus these lines are also guaranteed to host gapless surface states.
In other words, a non-zero net chirality guarantees that the planes $k_y = -\pi$ and $k_y = 0$ are Chern insulators.
To demonstrate this, we compute the dispersion of the surface states of the model in the main text along $k_y = -\pi$ and find the presence of one such state per surface (\cref{fig:figure2.1}).
This fact can then be leveraged to construct models with non-zero chirality on $K^2\times S^1$ by starting with Chern insulator models.

\section{The $\mathbb{Z}_2$ invariant on $\text{\emph{K}}^{\text{2}}$ and edge states}

\begin{figure*}[ht]
    \centerline{%
    \includegraphics[scale=1]{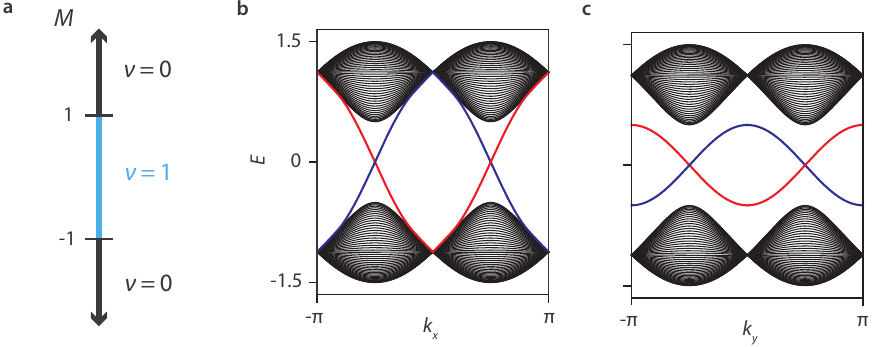}}
    \caption{%
        \textbf{a},~Phase diagram for two-dimensional tight-binding model in \cref{eq:d_functions_K2}.
        For $|M|>1$, the system is a trivial insulator, whereas for $|M|<1$ the system is a Klein bottle insulator.
        \textbf{b--c},~ Energy dispersion for $M = 0.5$ with open boundary conditions along $y$ (\textbf{b}) and $x$ (\textbf{c}).
        Both edges host topological states, but such states are pinned to the bulk bands along $y$, whereas they are unpinned but in-gap along $x$.
        }
    \label{fig:figure3}
\end{figure*}

For a two-dimensional toroidal momentum-space fundamental domain $T^2$, a $k_y$-dependent Berry phase $\gamma(k_y)$ can be obtained by integrating the Berry connection along non-contractible paths in the $k_x$ direction.
The two zone edges $k_y = -\pi$ and $k_y = \pi$ are physically equivalent and oriented in the same direction, leading to the constraint $\gamma(k_y = \pi) = \gamma(k_y = -\pi) \pmod{2\pi}$.
The set of curves obeying such a constraint can be labeled by an integer, the Chern number, which corresponds to the Berry phase winding number.

If the Bloch Hamiltonian obeys \cref{eq:H_cons}, then the topology of the fundamental domain is instead that of a Klein bottle $K^2$. As discussed in the main text, this leads to a $\mathbb{Z}_2$ classification on the Berry phase, associated with the invariant:
\begin{align}
    \nu = W_\pi \bmod 2,
    \label{eq:nu}
\end{align}
where $W_\pi$ counts the number of intersections of the Berry phase with the line $\gamma = \pi$. 

By the bulk--boundary correspondence, a non-trivial value of $\nu$ is expected to lead to topological edge states. 
Physically, the value of $k_y$ at which the Berry phase intersects $\gamma = \pi$ corresponds to a one-dimensional chain with a non-trivial charge polarization of $\pi$, leading to an in-gap state. As $k_y$ is varied away from this value, continuity of energy levels then gives rise to an in-gap edge band.

To explicitly show the existence of these edge states, we take the following two-dimensional model:
\begin{align}
    \begin{split}
    \label{eq:d_functions_K2}
    d_x(\mathbf{k}) &= \cos{k_x}, \\
    d_y(\mathbf{k}) &= \sin{k_x}\cos{k_y}, \\
    d_z(\mathbf{k}) &= M + \sin{k_x}\sin{k_y},
    \end{split}
\end{align}
where the mass term $M$ controls the topological phase transitions.
The phase diagram for this model is shown in \cref{fig:figure3}a.
For $|M| < 1$, $\nu = 1$ and we find topological states for both edges.
Note further that edge states on the $y$-edges are gapless but non-chiral (\cref{fig:figure3}b), whereas those on the $x$-edges are in-gap, featuring a M\"obius-like twist on the interval $-\pi \le k_y < 0$ (\cref{fig:figure3}c)~\cite{Shiozaki2015, Chen2022}.

\section[{Local calculation of $\mathbb{Z}_2$ charge: two-sided Klein bottle}]{Local calculation of $\mathbb{Z}_2$ charge: the two-sided Klein bottle}

\begin{figure*}[ht]
    \centerline{%
    \includegraphics[scale=1]{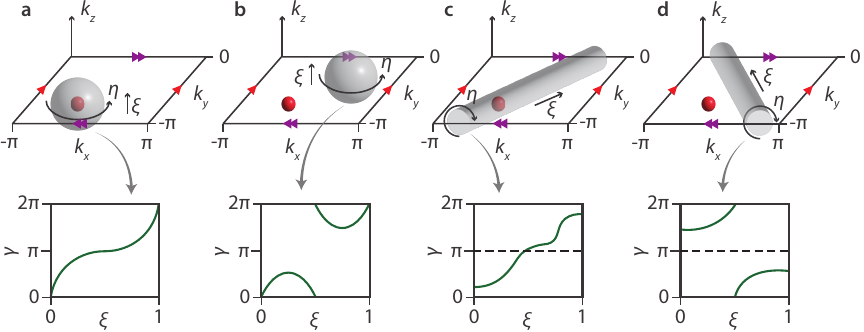}}
    \caption{%
        \textbf{a},~The Chern number of a Weyl point can be calculated by enclosing it within a sphere and integrating the Berry connection along the circumferential direction, $\eta$, to obtain the Berry phase $(\gamma)$, and plotting it as a function of the polar direction, $\xi$. The winding number of $\gamma(\xi)$ is the Chern number.
        \textbf{b},~When no Weyl points are present within the sphere, the Chern number is vanishing. 
        \textbf{c},~The invariant $\nu$ is sourced by Weyl points of an odd Chern number and can be calculated locally by enclosing the Weyl point within a two-sided Klein bottle. The Berry phase is calculated along the $\eta$ direction and plotted as a function of the extended direction, $\xi$. 
        \textbf{d},~When no Weyl points (or Weyl points of an even Chern number) are present inside the two-sided Klein bottle, the value of $\nu$ is vanishing.    
        }
    \label{fig:figure4}
\end{figure*}

In Fig.~3f of the main text, we evaluated the Klein bottle $\mathbb{Z}_2$ invariant $\nu(k_z)$ for various $k_z$ planes and showed that it changes as $k_z$ passes through an odd number of Weyl points. 
This suggests that $\nu$ is sourced by Weyl points and that a non-trivial value can be associated with odd Chern numbers.
To show that the Weyl points indeed carry this $\mathbb{Z}_2$ charge, we must perform a local calculation of $\nu$ around the Weyl points, similar to how the Chern number can be calculated from the net winding of Berry phases over polar loops on a sphere enclosing the Weyl point (\cref{fig:figure4}{a, b}). 
However, since $\nu$ is a topological invariant defined on the Klein bottle, this necessitates enclosing the Weyl point inside a Klein bottle. 
In orientable spaces, they are necessarily one-sided objects that do not enclose a volume. 
However, because the momentum-space manifold $K^2\times S^1$ is itself non-orientable, it is possible to embed Klein bottles within this space that have two sides and enclose a volume---these will be referred to as two-sided Klein bottles.
In other words, the properties of non-orientability and one-sidedness depend on embedding and need not imply each other~\cite{woll1971one}.

The two-sided Klein bottles are constructed such that they can be shrunk to a point along one direction but are extended along the other direction, as shown in \cref{fig:figure4}{c, d}.
We compute the Berry phase by integrating the Berry connection along the azimuthal direction, parametrized by an angular variable $\eta$, and plot it as a function of $\xi$, which parameterizes the extended direction. 
The value of $\nu$ can be found from $\gamma(\xi)$ for the cases when either a single Weyl point of unit charge is enclosed within the two-sided Klein bottle or no Weyl points are present. 
The non-trivial case with $\nu = 1$ occurs when the Weyl point is enclosed within the two-sided Klein bottle (\cref{fig:figure4}c). 
The trivial case results when no Weyl points are enclosed within, \ie $\nu=0$ (\cref{fig:figure4}d). 
This implies that $\nu$ is indeed associated with and sourced by Weyl points; accordingly, we associate a $\mathbb{Z}_2$ charge to each.

\begin{figure*}[ht]
    \centerline{%
    \includegraphics[scale=1]{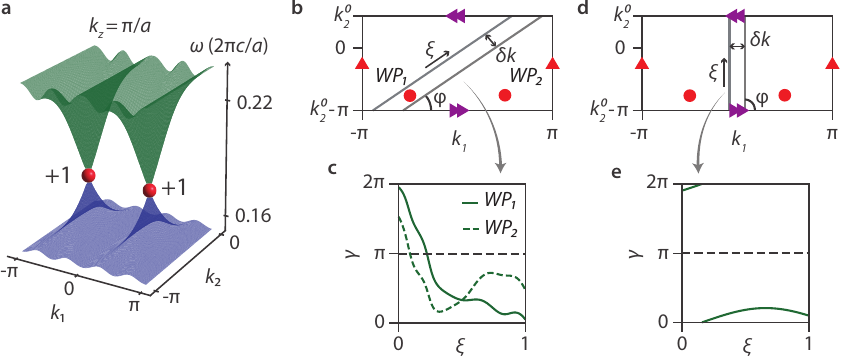}}
    \caption{%
        \textbf{a},~The lowest two bands of the photonic crystal system from the main text at $k_z = \pi/a$, as a function of $k_1$ and $k_2$. The two Weyl points, shown as red spheres, both carry a Chern number of $+1$.
        \textbf{b},~A two-dimensional slice showing a cross section of the two-sided Klein bottle (grey) enclosing one of the Weyl points. The two Weyl points are denoted as $WP_1$ and $WP_2$. The two-sided Klein bottle has thickness $\delta k$, tilt angle $\varphi$ (in radians) and its extended direction is parametrized by $\xi$.
        Throughout these calculations, we set $\delta k = 0.5$ and $k_2^0 = 1.07$.
        \textbf{c},~When the two-sided Klein bottle encloses either Weyl point, the Berry phase calculation yields $\nu = 1$. Tilt angles used for $WP_1$ and $WP_2$ are $\varphi = 0.6$ and $\varphi = 2.54$, respectively.
        \textbf{d},~A cross section of the two-sided Klein bottle enclosing no Weyl points.
        \textbf{e},~In this case, the Berry phase calculation yields $\nu = 0$. The tilt angle used here is $\varphi = \pi/2$.
        }
    \label{fig:figure4.5}
\end{figure*}

We now apply this procedure for the Weyl points that occur in the photonic crystal system in the main text (\cref{fig:figure4.5}a).
We note that, the glide symmetry in \cref{eq:H_cons} only involves a shift along $k_2$, and as a result, the in-plane region $-\pi \le k_1 < \pi$, $-\pi+k_2^0 \le k_2 < \pi+k_2^0, -\pi \le k_z < \pi$ forms a fundamental domain with the topology of $K^2\times S^1$, for any value of $k_2^0$.
For ease of computation, we choose a fundamental domain indexed by a non-zero $k_2^0$ in order to enclose the Weyl points with two-sided Klein bottles (\cref{fig:figure4.5}b).
Indeed, when the two-sided Klein bottle encloses either Weyl point, we obtain $\nu = 1$ (\cref{fig:figure4.5}c) and when no Weyl points are enclosed, we obtain $\nu = 0$ (\cref{fig:figure4.5}d, e). 

\section{Relation between $\mathbb{Z}_2$ charge and Chern number}

\begin{figure*}[ht]
    \centerline{%
    \includegraphics[scale=1]{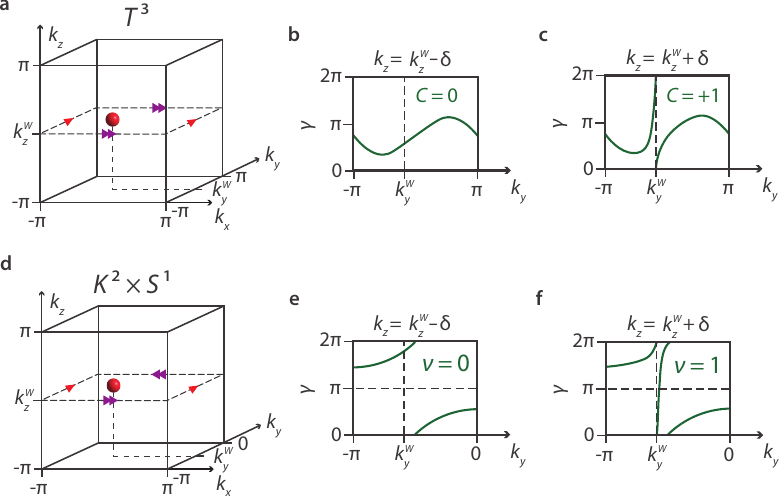}}
    \caption{%
    \textbf{a},~A Weyl point with $C=+1$ at $(k_y, k_z) = (k_y^W, k_z^W)$ on a toroidal momentum-space manifold. \textbf{b},~Berry phase at a $k_z = k_z^W - \delta$, for some small $\delta$. \textbf{c},~Berry phase at a $k_z = k_z^W + \delta$, showing that a jump in Berry phase by $2\pi$ at $k_y = k_y^W$ leads to a change in the Chern number. \textbf{d},~A Weyl point with $C=+1$ at $(k_y, k_z) = (k_y^W, k_z^W)$ on a $K^2 \times S^1$ manifold. \textbf{e},~Berry phase at a $k_z = k_z^W - \delta$ with $\nu = 0$. \textbf{f},~Berry phase at a $k_z = k_z^W + \delta$, showing that a jump in Berry phase by $2\pi$ at $k_y = k_y^W$ leads to a change in $\nu$ by unity.
        }
    \label{fig:figure5}
\end{figure*}

In this section, we discuss the relationship between the $\mathbb{Z}_2$ charge of Weyl points, $\nu$, and their $\mathbb{Z}$ charge, given by the Chern number, $C$. In particular, we show that $\nu$ and $C$ satisfy
\begin{align}
    \nu = C \bmod 2.
    \label{eq:nuisCmod2}
\end{align}
Consider a transition between a trivial insulator and a Chern insulator phase mediated by a Weyl point.
This could be thought of as occurring in a two-dimensional system, induced by a changing mass parameter, or in two-dimensional cuts of a three-dimensional system (\cref{fig:figure5}a).
The exact mechanism of such a transition can be made evident by examining the Berry phase, $\gamma$, of constant-$k_z$ cuts immediately before and after a Weyl point (\cref{fig:figure5}{b, c}). 
The Berry phase is an analytic function of $k_y$ everywhere except at the momentum of the Weyl point, $k_y^W$, where it becomes ill-defined. 
It can be shown that on passing through a Weyl point, the value of the Berry phase jumps by $2\pi C$, where $C$ is the Chern number of the Weyl point~\cite{Yoshida.1:2023, Yoshida.2:2023}.
As a result, the winding number of $\gamma(k_y)$ is increased by $C$.
Similarly, when the fundamental domain is reduced to $K^2\times S^1$, the same jump in Berry phase due to a Weyl point leads to an increase in $W_{\pi}$ by $C$ (\cref{fig:figure5}{e, f}), which implies \cref{eq:nuisCmod2} \cf \cref{eq:nu}.

\section{Inversion and time-reversal symmetries}

\begin{figure}[ht]
    \centerline{%
    \includegraphics[scale=1]{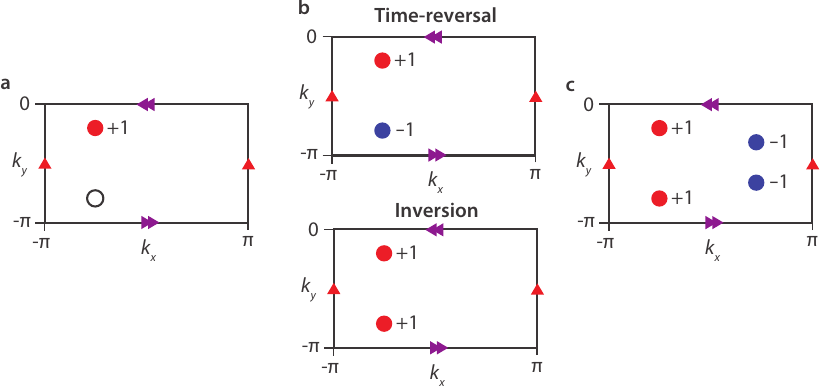}}
    \caption{%
        \textbf{a}, Generic Weyl point distribution under either time-reversal or inversion symmetry in the minimal case. 
        \textbf{b}, Time-reversal/inversion requires that the symmetry-related Weyl point have the opposite/same chirality as the original (here, with chiralities $\mp 1$).
        \textbf{c}, An example of an inversion-symmetric Weyl point configuration that does not circumvent the Nielsen--Ninomiya theorem.        
        }
    \label{fig:figure12}
\end{figure}

In a system with a $K^2 \times S^1$ fundamental domain, the presence of inversion or time-reversal symmetry can affect whether or not the Nielsen--Ninomiya theorem is circumvented, as we detail below.
 
Let us illustrate this in the minimal case of two Weyl points on the fundamental domain.
Under either time-reversal or inversion symmetry, a Weyl point at $(k_x, k_y, k_z)$ is necessarily accompanied by another one on the fundamental domain, located at $(k_x, -k_y-\pi, -k_z)$ via composing either of these symmetries with the glide symmetry
(\cref{fig:figure12}a).
If the system is inversion-symmetric, then the two Weyl points must be of the same chirality, as both the glide symmetry and inversion flip the chirality; if the system is time-reversal-symmetric, then the two Weyl points have opposite chirality, since time-reversal does not flip its sign (\cref{fig:figure12}b).
Thus, the Nielsen--Ninomiya theorem must be circumvented in inversion-symmetric systems, and it must be preserved if time-reversal is present.
We note that these conclusions are only necessarily true for our choice of the fundamental domain and if the number of Weyl points in the fundamental domain is an odd multiple of two.
For example, with four Weyl points on the fundamental domain, the Nielsen--Ninomiya theorem may remain applicable even in the presence of inversion (\cref{fig:figure12}c).

\color{black}
\section{Proof of $\mathbb{Z}_2$ no-go theorem via Fermi arcs}

\begin{figure*}
    \centerline{%
    \includegraphics[scale=1]{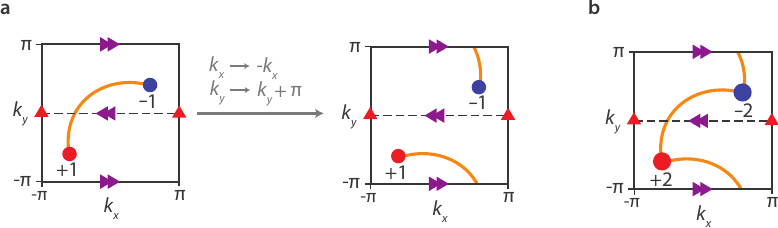}}
    \caption{%
        \textbf{a},~Projections of Weyl points on the torus for a model with a single Weyl point in $K^2 \times S^1$. On truncating the system along the $z$ direction, there must be a Fermi arc connecting the two Weyl points.
        Such a Fermi arc cannot conform by itself to the momentum-space glide symmetry, and therefore a single Fermi arc is disallowed.
        \textbf{b},~Projections of Weyl points on the torus for a model with two Weyl point in $K^2 \times S^1$, here taken to have the same in-plane coordinates.
        The additional Fermi arc then allows the system to be fully symmetric under the glide symmetry.
        }
    \label{fig:figure7}
\end{figure*}

We now give an alternative proof that there must always be an even number of singly charged Weyl points on $K^2\times S^1$.
Consider the simple case of a single Weyl point in the $K^2\times S^1$ fundamental domain spanned by $-\pi \le k_x, k_z < \pi$ and $-\pi \le k_y < 0$. This configuration leads to a second Weyl point of opposite chirality on the other half of the torus, \ie in the domain spanned by $-\pi \le k_x, k_z < \pi$ and $0 \le k_y < \pi$, such that the Nielsen--Ninomiya theorem holds on $T^3$ (\cref{fig:figure7}a).
As a result, the $(k_x, k_y)$ surface Brillouin zone must host one Fermi arc per surface connecting them.
The surface Hamiltonian is also constrained to obey the momentum-space glide symmetry, and therefore so are the Fermi arcs.
However, there is \emph{no} Fermi arc configuration that is invariant under this symmetry; instead, there must be at least two pairs of Weyl points on $T^3$---and therefore a pair of Weyl points on $K^2\times S^1$. In this case, the additional Fermi arcs are mapped into the previous ones under the symmetry, and vice-versa (\cref{fig:figure7}b).

\section{Nielsen--Ninomiya: many-band case}

Here, we briefly discuss the case of many connected bands and the Nielsen--Ninomiya theorem.
We consider a system with $N$ bands that host finitely many Weyl points between successive pairs of bands such that they are all connected.
For a subset of bands $m<N$, the Hamiltonian projected into this subspace, $H_m(\mathbf{k})$, is well-defined everywhere except at the Weyl points where the bands touch.
In a spherical neighborhood around each such Weyl point, an integer-valued winding number can be assigned using $H_m(\mathbf{k})$, similar to the index of a vector field around one of its zeros.
On a toroidal manifold, these winding numbers sum to zero and as a result, the total chirality of the Weyl points vanishes.
See e.g., Ref.~\citenum{witten2016three} for additional details.
Similarly, on the fundamental domain $K^2\times S^1$ in our models, the discontinuity argument of $\mathbf{d}(\mathbf{k})$ through the Poincar\'e--Hopf theorem can be generalized to the many-band case by invoking the notion of a Grassmanian manifold to define indices of a vector field~\cite{Milnor1974}.

\begin{figure}[h]
    \centerline{%
    \includegraphics[scale=1]{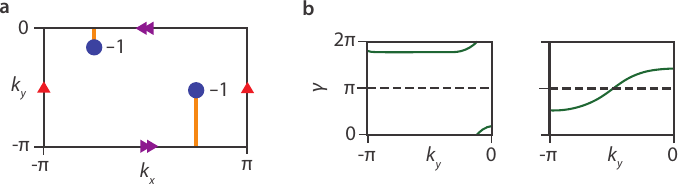}}
    \caption{%
        \textbf{a},~Fermi arcs for the bottom two bands, obtained on truncating the system along the $z$ direction for the model in~\cref{eq:four_band}, connecting projections of Weyl points with charges of the same sign via an orientation-reversing path.
        \textbf{b},~Berry phase, $\gamma$, calculated on $K^2$ for $k_z$ cuts below (left) and above (right) a Weyl point, indicating that $\nu(k_z)$ shifts by unity upon sweeping across a Weyl point.
        }
    \label{fig:figure10}
\end{figure}

We now illustrate our theory using a four-band model:
\begin{align}
    H(\mathbf{k}) = d_0(\mathbf{k}) \Gamma_{10} + d_x(\mathbf{k}) \Gamma_{11} + d_y(\mathbf{k}) \Gamma_{12} + d_z(\mathbf{k}) \Gamma_{13}, 
\end{align}
where $\Gamma_{ij} = \sigma_i \otimes \sigma_j$.
We can choose the following $\mathbf{d}$ components:
\begin{align}
    \begin{split}
        d_0(\mathbf{k}) &= 1, \\
        d_x(\mathbf{k}) &= \cos{k_x}, \\
        d_y(\mathbf{k}) &= \sin{k_z} - \sin{k_x}\sin{k_y}  - \tfrac{1}{2}, \\
        d_z(\mathbf{k}) &= \cos{k_z} + \sin{k_x}\cos{k_y} + 1.
    \end{split}
    \label{eq:four_band}
\end{align}
This model exhibits two Weyl points between both the lowest two bands and highest two bands, and all Weyl points on the fundamental domain have Chern number $-1$ (\cref{fig:figure10}a), leading to a circumvention of the Nielsen--Ninomiya theorem for both pairs of bands.
As shown in~\cref{fig:figure10}b, the $\mathbb{Z}_2$ invariant changes by unity upon crossing an odd number of Weyl points, as expected.

\begin{figure*}[ht]
    \centerline{%
    \includegraphics[scale=1]{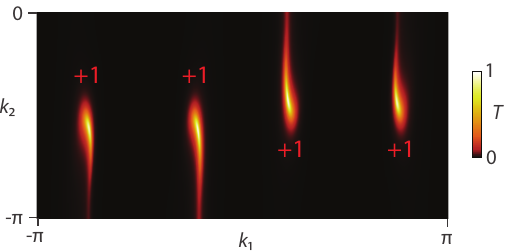}}
    \caption{Transmission spectrum of the Fermi arcs of the photonic crystal system between bands 2 and 3. These Fermi arcs reside on the top surface of the finite-in-$z$ system on the $K^2$ surface Brillouin zone formed in $(k_1, k_2)$ space between bands 2 and 3.
    }
    \label{fig:figure8}
\end{figure*}

The case of many bands is relevant for the photonic crystal system, as isolated pairs of bands hosting Weyl points do not readily occur.
However, the applicability (or the circumvention in the case of $K^2 \times S^1$) of the Nielsen--Ninomiya theorem holds for an arbitrary set of bands as argued above. 
In Fig. 3 of the main text, we showed that bands 1 and 2 of the photonic crystal host two Weyl points of the same relative chirality. 
Similarly, bands 2 and 3 host four Weyl points of the same chirality with Fermi arcs connecting them via orientation-reversing paths (\cref{fig:figure8}).

\section{Methods} 
For the fabrication of the 1D photonic crystals (PhCs), we employ plasma-enhanced chemical vapor deposition (PECVD) using the SAMCO PD-220 system to deposit a stack of silicon (Si) and silica (SiO$_2$) layers onto a fused silica substrate (microscope coverslips of thickness \SI{180}{$\mu$ m}). 
The deposition of Si is achieved from Ar and SiH$_4$ precursor gases at \SI{350}{\degree{C}} with gas flow rates of \SI{28}{sccm} and \SI{500}{sccm} respectively. The deposition of SiO$_2$ is achieved from N$_2$O and SiH$_4$ precursor gases at \SI{350}{\degree{C}} with gas flow rates of \SI{460}{sccm} and \SI{5}{sccm} respectively. 
These parameters lead to a deposition rate of approximately \SI{55}{nm/min} for both materials.
The deposition time is used to control the thicknesses of the individual layers.
We set the lattice constant $a = \SI{200}{nm}$ and fabricate six unit cells, \ie 24 layers, and use three additional layers as cladding, each with thickness $a/2$. 
These cladding layers increase the lifetime of the surface states, allowing for their observation in the transmission spectrum of the 1D PhCs.
We find that any fabrication errors that may be present are not large enough to cause any meaningful deviation of the observed spectral features compared to simulations.

To characterize the 1D PhCs, we measure their transmission spectrum at normal incidence using a Cary 5000 UV-Vis-NIR spectrophotometer in the wavelength range of \SI{650}{nm} to \SI{1400}{nm}. 
The transmitted spectrum is always shown normalized to that of the bare substrate.

The transmission spectra of the 1D PhCs shown in Figures 3c,d,e of the main text are calculated using the Rigorous Coupled Wave Analysis (RCWA) method, as implemented in \textsc{Stanford Stratified Structure Solver} (S\textsuperscript{4})~\cite{StanfordS4} and the calculation of the photonic band structure shown in Figure 3b of the main text is performed using the \textsc{MIT Photonic Bands} package~\cite{MPB}.

We note that for the these PhCs, a simple transfer matrix approach is sufficient to calculate the transmission spectra which we describe here.
Each layer $i$ in the 1D PhC is characterized by a thickness $t_i$ and a refractive index $n_i$. 
The characteristic matrix relating the input and output electric fields through this layer, at normal incidence, is given by
\begin{align}
        M_i(\lambda) = \begin{pmatrix}
            \cos z & {i}{n_i^{-1}}\sin z\\
            in_i\sin z & \cos z
        \end{pmatrix},
\end{align}
where $\lambda$ is the operating wavelength of light and $z=2\pi n_i t_i / \lambda$. The characteristic matrix for the 1D PhCs is the product of the characteristic matrices of each layer, \ie
\begin{align}
        \tilde{M}(\lambda) = \prod_i M_i(\lambda) = 
        \begin{pmatrix}
        \tilde{M}_{11}(\lambda) & \tilde{M}_{12}(\lambda) \\
        \tilde{M}_{21}(\lambda) & \tilde{M}_{22}(\lambda)
        \end{pmatrix}.
\end{align}
The transmission as a function of wavelength through the multilayer structure is then
\begin{align}
        T(\lambda) = 1 - \left| \frac{\tilde{M}_{21}(\lambda)}{\tilde{M}_{11}(\lambda)} \right|^2.
\end{align}

\FloatBarrier
\bibliographystyle{apsrev4-2-longbib}
\bibliography{bibliography_supp}